\documentclass[aps,prd,superscriptaddress,nofootinbib,amsmath,amssymb,reprint,floatfix]{revtex4-2}
\usepackage[T1]{fontenc}
\usepackage{lmodern} 
\usepackage{bm}  
\usepackage{booktabs}
\usepackage{graphicx}

\usepackage[colorlinks=true,citecolor=blue,linkcolor=blue,urlcolor=blue]{hyperref}

\newcommand{\LCDM}{$\Lambda$CDM}
\newcommand{\betastar}{\beta_{\ast}}

\begin{document}

\title{Phenomenological Constraints on a Quadratically Deformed FLRW Minisuperspace}

\author{Gamal G. L. Nashed}
\email{nashed@bue.edu.eg}
\affiliation{Centre for Theoretical Physics, The British University in Egypt, El Sherouk City, Cairo 11837, Egypt}
\affiliation{Centre for Space Research, North-West University, Potchefstroom 2520, South Africa}

\date{\today}

\begin{abstract}
We study a quadratic deformation of the reduced Poisson algebra in a homogeneous FLRW minisuperspace. Although motivated by generalized uncertainty principle models, this construction is not a covariant theory of quantum gravity; its dimensionless coefficient is therefore treated only as an effective parameter of the chosen minisuperspace and fiducial-volume convention. At first order, the deformation modifies the homogeneous Friedmann equation. We constrain the model using 31 cosmic-chronometer measurements, 1580 uncalibrated Pantheon$+$ Hubble-flow light-curve entries, and the 13-component DESI DR2 baryon-acoustic-oscillation likelihood, treating the sound horizon as a nuisance parameter. For the baseline normalization, we obtain $\betastar=-0.086^{+0.040}_{-0.032}$ at $68\%$ credibility and $-0.142<\betastar<0.005$ at $95\%$ credibility, so the undeformed limit remains consistent with the data. The corresponding cosmographic parameters are $q_0=-0.416^{+0.070}_{-0.069}$, $j_0=0.162^{+0.445}_{-0.461}$, and $z_{\mathrm{tr}}=0.688^{+0.027}_{-0.026}$. Compared with flat \LCDM{}, the model reduces the minimum chi-square by $4.687$, but the AIC preference is weak and the BIC favors \LCDM{}. The results also depend on higher-order normalization choices and the perturbative cutoff: with strict order-by-order normalization and $\eta_{\max}^{\rm strict}<0.05$, the $95\%$ interval becomes $-0.099<\betastar<0.013$. Overall, the data provide a conditional null constraint on this phenomenological minisuperspace model and do not support a fundamental minimum length or an observable quantum-gravity effect.
\end{abstract}

\maketitle

\section{Introduction}

Quantum-gravity models often suggest that the Heisenberg uncertainty relation may be modified near a fundamental length scale \cite{Veneziano1986,Amati1989,Maggiore1993,Kempf1995,Garay1995}. A common example is the quadratic generalized uncertainty principle (GUP), which introduces momentum-dependent corrections to the canonical algebra \cite{Kempf1995,Ali2009,Hossenfelder2013}. However, such a modification does not automatically produce observable effects at low energies. Effective-field-theory arguments and the decoupling principle imply that physics at the Planck scale should be strongly suppressed at much lower energy and curvature scales \cite{AppelquistCarazzone1975,Donoghue1994}. Since $H_0/M_{\rm Pl}\sim10^{-61}$, any correction directly controlled by the present cosmological curvature would be extremely small. An observable late-time quantum-gravity effect would therefore require a clear mechanism that avoids this suppression.

One way to study GUP-inspired corrections in cosmology is through a homogeneous FLRW minisuperspace model \cite{Battisti2009}. This approach preserves the Hamiltonian constraint and recovers standard cosmology in the undeformed limit. However, it is not derived from a local four-dimensional theory and does not explain how Planck-scale effects could evade low-energy decoupling. Because local gravitational degrees of freedom have been removed, the resulting dynamics may depend on the choice of canonical variables, fiducial cell, and perturbative prescription. We therefore interpret the deformation parameter studied here as an effective coefficient of a specific reduced Hamiltonian model, rather than as a fundamental quantum-gravity parameter.

The deformed minisuperspace model also determines only the background expansion history. It does not provide a covariant action or unique equations for scalar, vector, and tensor perturbations. For this reason, inserting the modified Hubble function into standard Einstein--Boltzmann equations would require additional assumptions not implied by the deformed algebra. We therefore restrict our analysis to late-time background observations and do not include CMB anisotropies, weak lensing, redshift-space distortions, or matter-power-spectrum data.

Our goal is to determine how strongly late-time background data constrain this reduced-phase-space deformation and whether any apparent preference is robust against normalization and truncation choices. We use cosmic chronometers (CC), which provide direct measurements of $H(z)$ \cite{JimenezLoeb2002,Moresco2016}; the Pantheon$+$ Type Ia supernova sample, which constrains relative luminosity distances \cite{Brout2022}; and the DESI DR2 baryon-acoustic-oscillation (BAO) likelihood \cite{DESIDR2}. Since the model does not predict pre-recombination perturbation physics, we treat the sound horizon at the drag epoch as an independent nuisance parameter.

We derive the modified Friedmann equation and examine two normalization prescriptions: an exact algebraic treatment of the truncated equation and a strict order-by-order expansion. We identify the parameter controlling the first-order approximation and test the sensitivity of the results to its cutoff. Using identical data and likelihood assumptions, we compare the GUP model with flat \LCDM{} and the Chevallier--Polarski--Linder (CPL) parametrization \cite{Chevallier2001,Linder2003}. We also calculate the deceleration parameter, jerk parameter, transition redshift, and effective background quantities. This allows us to assess whether any improvement in the fit remains within the perturbative regime of the model.

The paper is organized as follows. Section~\ref{sec:model} presents the GUP-modified Friedmann equation, its normalization choices, higher-order structure, and effective-background interpretation. Section~\ref{sec:observables} introduces the background observables, and Sec.~\ref{sec:data} describes the data, priors, and likelihoods. Section~\ref{sec:comparison} presents the reference \LCDM{} and CPL models. Results and statistical comparisons are given in Sec.~\ref{sec:results}, while Secs.~\ref{sec:robustness} and \ref{sec:cosmography} examine perturbative robustness and cosmographic constraints. Section~\ref{sec:conclusions} summarizes our conclusions.

\section{Phenomenologically deformed minisuperspace background}\label{sec:model}

We consider a spatially flat FLRW line element,
\begin{equation}
 ds^2=-N^2(t)dt^2+a^2(t)d\bm{x}^2,
\end{equation}
and choose the cosmic-time gauge $N=1$ after variation.  For a unit
fiducial comoving cell, the undeformed reduced action, after elimination of separately conserved matter variables, can be written as
\begin{equation}
 S=\int dt\left[-\frac{3a\dot a^{\,2}}{8\pi G N}-Na^3\rho\right].
 \label{eq:reduced_action}
\end{equation}
In this undeformed system, the momentum conjugate to the scale factor is
\begin{equation}
 p_a=\frac{\partial L}{\partial\dot a}
 =-\frac{3a\dot a}{4\pi G N}.
 \label{eq:canonical_momentum}
\end{equation}
The Legendre transformation gives $\mathcal H_{\rm tot}=N\mathcal H$,
where the lapse is a Lagrange multiplier.  Variation with respect to
$N$ gives the Hamiltonian constraint \cite{MTW1973}
\begin{equation}
 \mathcal{H}=-\frac{2\pi G}{3}\frac{p_a^2}{a}+a^3\rho=0,
 \label{eq:Hamiltonian}
\end{equation}
so that
\begin{equation}
 p_a^2=\frac{3}{2\pi G}a^4\rho.
 \label{eq:pa}
\end{equation}
We now retain the function $\mathcal H(a,p_a)$ in
Eq.~\eqref{eq:Hamiltonian} and deform its symplectic structure. This is
a new Hamiltonian prescription, rather than a second variation of
Eq.~\eqref{eq:reduced_action}. In particular,
Eq.~\eqref{eq:canonical_momentum} is the undeformed momentum--velocity
relation and must not also be imposed on the deformed dynamics.
The quadratic minisuperspace deformation is
\begin{equation}
 \{a,p_a\}=1+\beta p_a^2,
 \label{eq:bracket}
\end{equation}
where the dimensionful coefficient $\beta$ satisfies $[\beta]=[p_a]^{-2}$.
The standard minimum-length interpretation of this quadratic algebra takes
$\beta>0$ \cite{Kempf1995}.  Here both signs are admitted in order to test
the resulting background phenomenologically.  Since
$\betastar=3\beta\rho_{c0}/(\pi G)$, $\betastar$ and $\beta$ have the same
sign.  For $\beta<0$, the factor $1+\beta p_a^2$ would vanish at
$p_a^2=|\beta|^{-1}$ if the bracket were extrapolated beyond its
perturbative domain.  Our inference remains restricted to
$|\beta p_a^2|<0.1$ and never approaches that point.  Nevertheless, this
algebraic distinction means that a negative fitted value must not be
identified with the conventional positive-$\beta$ minimum-length branch.
The prescription has a local first-order action on the domain
$1+\beta p_a^2>0$:
\begin{align}
 S_\beta&=\int dt\,[P(p_a)\dot a-N\mathcal H(a,p_a)],
 \label{eq:deformed_action}\\
 P(p_a)&=\int_0^{p_a}\frac{du}{1+\beta u^2}
 =p_a-\frac{\beta p_a^3}{3}+\mathcal O(\beta^2).
 \label{eq:Darboux_momentum}
\end{align}
Here $(a,P)$ is a canonical pair, whereas $(a,p_a)$ obeys
Eq.~\eqref{eq:bracket}. Varying $p_a$, $a$, and $N$ gives
\begin{align}
 \dot a&=N(1+\beta p_a^2)\partial_{p_a}\mathcal H,\\
 \dot p_a&=-N(1+\beta p_a^2)\partial_a\mathcal H,
 \qquad \mathcal H=0.
\end{align}
Thus the constraint is preserved, $\dot{\mathcal H}=0$, and the
deformed momentum--velocity relation follows from this action.
This reduced action does not supply a four-dimensional covariant
completion or equations for inhomogeneous perturbations. After setting
$N=1$, Hamilton's equation gives
\begin{equation}
 \dot a=\left(1+\beta p_a^2\right)
 \frac{\partial\mathcal{H}}{\partial p_a}
 =-\left(1+\beta p_a^2\right)\frac{4\pi G}{3}\frac{p_a}{a}.
\end{equation}
Keeping terms only through first order in $\beta$ and using Eq.~\eqref{eq:pa}, we obtain
\begin{equation}
 H^2=\frac{8\pi G}{3}\rho
 \left(1+\frac{3\beta}{\pi G}a^4\rho\right)
 +\mathcal{O}(\beta^2).
 \label{eq:FriedmannDimensional}
\end{equation}

We normalize $a_0=1$ and define
\begin{equation}
 \rho_{c0}=\frac{3H_0^2}{8\pi G},
 \qquad
 \betastar=\frac{3\beta\rho_{c0}}{\pi G}.
 \label{eq:betastar}
\end{equation}

\subsection{Status of the deformation and low-energy decoupling}
\label{sec:decoupling}

Equation~\eqref{eq:bracket} is an ansatz imposed after reduction to the
homogeneous phase space.  It is not derived by integrating out a specified
set of ultraviolet fields, and no covariant action is known here whose
constraint algebra reduces uniquely to Eq.~\eqref{eq:bracket}.  The
calculation should therefore be distinguished from a gravitational
effective field theory in which higher-curvature operators and their
suppression scales are fixed before the cosmological reduction.

This distinction is essential at late times.  If a correction is produced
by local physics at a mass scale $M_{\rm UV}$, decoupling generically yields
a dimensionless background correction of the schematic form
\begin{equation}
 \frac{\delta H^2}{H^2}
 \sim c_p\left(\frac{H}{M_{\rm UV}}\right)^p,
 \qquad p>0,
 \label{eq:decoupling_estimate}
\end{equation}
up to model-dependent powers of other infrared scales.  For
$M_{\rm UV}\sim M_{\rm Pl}$ and $H\sim H_0$, this is negligible unless a
separate enhancement or nondecoupling mechanism is demonstrated.  The
minisuperspace bracket in Eq.~\eqref{eq:bracket} does not supply such a
mechanism, and we do not interpret the fitted value of $\betastar$ as an
exception to Eq.~\eqref{eq:decoupling_estimate}.

There is also no convention-independent map from $\betastar$ to a minimum
length.  Restoring a fiducial comoving volume $V_0$ in the reduced action
rescales the canonical momentum, $p_a\propto V_0$, and the coefficient
multiplying $p_a^2$ must be rescaled correspondingly if the same reduced
equation is to be retained.  A posterior for $\betastar$ obtained after
fixing $V_0=1$ is thus a constraint on that conventionally normalized
reduced model.  It is neither a bound on a unique microscopic coupling nor
a measurement of the Planck scale.  The use of the label ``GUP'' below
identifies the algebraic motivation for the ansatz only.

For separately conserved radiation, nonrelativistic matter, and a cosmological constant, introduce
\begin{equation}
 X(a)=\Omega_{r0}a^{-4}+\Omega_{m0}a^{-3}+\Omega_{\Lambda0}.
\end{equation}
The expansion law becomes
\begin{equation}
 E^2(a)\equiv\frac{H^2(a)}{H_0^2}
 =X(a)+\betastar a^4X^2(a).
 \label{eq:E2}
\end{equation}

\subsection{Formal structure beyond first order}

The order counting can be displayed before imposing the observational
truncation.  If Eq.~\eqref{eq:bracket} is treated as an exact
phenomenological bracket, Hamilton's equation and the constraint give
\begin{equation}
 H^2=\frac{8\pi G}{3}\rho\left(1+\beta p_a^2\right)^2.
 \label{eq:formal_exact_dimensional}
\end{equation}
Using $\beta p_a^2=(\betastar/2)a^4X(a)$, the formal dimensionless
expression is
\begin{align}
 E_{\rm formal}^2(a)
 &=X(a)\left[1+\frac{\betastar}{2}a^4X(a)\right]^2 \nonumber\\
 &=X(a)+\betastar a^4X^2(a)
 +\frac{\betastar^2}{4}a^8X^3(a).
 \label{eq:formal_exact}
\end{align}
Equation~\eqref{eq:E2} is the first-order truncation.  The last term
exhibits the first omitted contribution and shows directly why
$\frac12\betastar a^4X$ controls the approximation.

Equation~\eqref{eq:formal_exact} is not used here as an all-orders
quantum-gravity prediction.  The quadratic bracket is itself an
effective prescription, and higher powers may receive contributions
from terms absent from Eq.~\eqref{eq:bracket}.  Retaining the square
would also change the present-day normalization to
\begin{equation}
 1=X_0\left(1+\frac{\betastar}{2}X_0\right)^2,
 \label{eq:formal_exact_closure}
\end{equation}
rather than Eq.~\eqref{eq:closure} below.  An untruncated analysis
therefore requires a separate inference and cannot be obtained by
inserting the $\betastar^2$ term into the already normalized
first-order fit.

\subsection{Present-day normalization}
Because $E(1)=1$ by definition, the density parameters and $\betastar$ obey
\begin{equation}
 1=X_0+\betastar X_0^2,
 \qquad
 X_0=\Omega_{r0}+\Omega_{m0}+\Omega_{\Lambda0}.
 \label{eq:closure}
\end{equation}
Thus the usual relation $X_0=1$ cannot simultaneously be imposed when $\betastar\neq0$.  Solving Eq.~\eqref{eq:closure} on the branch continuous at $\betastar=0$ gives\footnote{Equation~(\ref{eq:X0}) corresponds to the solution that is
continuous in the undeformed limit, satisfying
$\lim_{\beta^{*}\to0}X_{0}=1$ and thereby recovering the standard
flat-FLRW normalization. The second algebraic root,
$X_{0}=(-1-\sqrt{1+4\beta^{*}})/(2\beta^{*})$, diverges as
$\beta^{*}\to0$ and therefore is not continuously connected to the
$\Lambda$CDM background. Since the present construction is obtained as
a perturbative deformation about the standard cosmology, only the branch
continuously connected to $\Lambda$CDM is retained in the baseline
exact-root prescription.}
\begin{equation}
 X_0=\frac{2}{1+\sqrt{1+4\betastar}},\qquad \betastar\geq-\frac14.
 \label{eq:X0}
\end{equation}
In the numerical implementation, $\Omega_{\Lambda0}$ is therefore a derived quantity,
\begin{equation}
 \Omega_{\Lambda0}=X_0-\Omega_{m0}-\Omega_{r0}.
 \label{eq:OmegaLambda}
\end{equation}
The reality of Eq.~\eqref{eq:X0} requires $\betastar\geq-1/4$, and the physical prior must also ensure $E^2(z)>0$ throughout the fitted redshift interval.  Equation~\eqref{eq:X0} solves the closure relation of the truncated phenomenological equation exactly.  Consequently, its dependence on $\betastar$ contains terms beyond first order; it is an operational normalization prescription for the baseline model, not a unique consequence of a strictly first-order expansion.

For comparison, a normalization performed consistently order by order may be written by defining
\begin{equation}
 Y(a)=\Omega_{r0}a^{-4}+\Omega_{m0}a^{-3}
      +1-\Omega_{r0}-\Omega_{m0},
 \qquad Y(1)=1.
\end{equation}
The first-order closure gives $X(a)=Y(a)-\betastar+\mathcal{O}(\betastar^2)$.  Substitution into Eq.~\eqref{eq:E2}, followed by a consistent removal of terms of order $\betastar^2$ and higher, yields
\begin{equation}
 E_{\rm strict}^{\,2}(a)
 =Y(a)+\betastar\left[a^4Y^2(a)-1\right]
 +\mathcal{O}(\betastar^2).
 \label{eq:E2_strict}
\end{equation}
This alternative satisfies $E_{\rm strict}(1)=1$ exactly at the retained order.  We keep Eq.~\eqref{eq:X0} as the baseline prescription and analyze Eq.~\eqref{eq:E2_strict} with a separate chain as a normalization-sensitivity test; samples from the two definitions are not mixed.

\subsection{Magnitude and redshift dependence of the correction}

The fractional correction to the undeformed expression is
\begin{equation}
 \epsilon_{\mathrm{GUP}}(a)
 \equiv\frac{E^2(a)-X(a)}{X(a)}
 =\betastar a^4X(a).
 \label{eq:fractional}
\end{equation}
During ideal radiation domination, $X(a)\simeq\Omega_{r0}a^{-4}$ and hence
\begin{equation}
 \epsilon_{\mathrm{GUP}}\simeq\betastar\Omega_{r0},
\end{equation}
which is constant.  During matter domination,
\begin{equation}
 \epsilon_{\mathrm{GUP}}\simeq\betastar\Omega_{m0}a,
\end{equation}
which decreases toward the past.  The model therefore does not, by itself, imply a fractional correction that grows without bound toward early times.

The truncation leading to Eq.~\eqref{eq:FriedmannDimensional} requires
\begin{equation}
 |\beta p_a^2|=\frac12|\betastar|a^4X(a)\ll1.
 \label{eq:validity}
\end{equation}
This condition must be checked at every sampled point and over the entire redshift range used in the likelihood.

\subsection{Normalization caveat}

As emphasized above, the numerical value of a minisuperspace momentum
depends on the normalization of the scale factor and on the fiducial
comoving cell used to reduce the spatial integral.  Equation~\eqref{eq:Hamiltonian}
and the inferred value of $\beta$ are therefore meaningful only after that
convention is fixed.  Throughout this work we take $a_0=1$ and a unit
fiducial comoving volume.  Constraints on $\betastar$ apply only within
this convention and cannot be translated into a model-independent
minimum-length or Planck-scale bound.

\subsection{Effective-background description}

It is useful to isolate the additive correction
\begin{align}
 \Delta_{\rm GUP}(z)&=\betastar(1+z)^{-4}X^2(z),\nonumber\\
 E^2(z)&=X(z)+\Delta_{\rm GUP}(z).
 \label{eq:delta_gup}
\end{align}
This decomposition permits a kinematic effective equation-of-state
parameter,
\begin{align}
 w_{\rm GUP}^{\rm eff}(z)
 &=-1+\frac{1+z}{3}
 \frac{d\ln|\Delta_{\rm GUP}(z)|}{dz}\nonumber\\
 &=-\frac73+\frac{2(1+z)}{3}\frac{X'(z)}{X(z)}.
 \label{eq:effective_w}
\end{align}
The limiting values are $w_{\rm GUP}^{\rm eff}\to1/3$ in ideal radiation
domination, $w_{\rm GUP}^{\rm eff}\to-1/3$ in matter domination, and
$w_{\rm GUP}^{\rm eff}\to-7/3$ when the constant term dominates.  These
limits describe the redshift scaling of the additive term; they do not
turn it into a separately conserved material fluid.  In particular, the
posterior preference for $\betastar<0$ corresponds to a negative additive
contribution, so the usual interpretation of a positive fluid density is
not applicable.

\section{Late-time background observables}\label{sec:observables}

In terms of redshift, $a=(1+z)^{-1}$, and Eq.~\eqref{eq:E2} defines
\begin{equation}
 H(z)=H_0E(z).
\end{equation}
For a spatially flat background, the comoving distance and luminosity distance are
\begin{align}
 D_M(z)&=c\int_0^z\frac{dz'}{H(z')},\\
 D_L(z)&=(1+z)D_M(z).
\end{align}
For a supernova at redshift $z$, the distance-dependent part of the
predicted apparent magnitude is
\begin{equation}
 5\log_{10}\!\left[\frac{D_L(z)}{\mathrm{Mpc}}\right]+25.
\end{equation}

BAO measurements constrain combinations of
\begin{equation}
 D_H(z)=\frac{c}{H(z)},
 \qquad
 D_V(z)=\left[zD_M^2(z)D_H(z)\right]^{1/3},
\end{equation}
usually divided by the sound horizon $r_d$ at the baryon drag epoch.  Since the present minisuperspace model does not determine a complete pre-recombination perturbation theory, our baseline late-time analysis treats $r_d$ as an independent nuisance parameter.  A value derived from a standard early-Universe Boltzmann calculation must not be presented as a GUP prediction.

\section{Data, likelihoods, and priors}\label{sec:data}

\begin{table*}[!t]
\centering
\caption{Public-release inputs. SHA-256 hashes of the exact local files used in the analysis are provided in the accompanying package.}
\label{tab:input_files}
\small
\begin{tabular}{ll}
\hline\hline
Release & Exact filename\\\hline
Pantheon$+$ & \texttt{Pantheon+SH0ES.dat}\\
Pantheon$+$ & \texttt{Pantheon+SH0ES\_STAT+SYS.cov}\\
DESI DR2 / Cobaya & \texttt{desi\_gaussian\_bao\_ALL\_GCcomb\_mean.txt}\\
DESI DR2 / Cobaya & \texttt{desi\_gaussian\_bao\_ALL\_GCcomb\_cov.txt}\\
\hline\hline
\end{tabular}
\end{table*}

The baseline likelihood combines 31 CC measurements treated with a diagonal error model
\cite{JimenezLoeb2002,Moresco2016}, the Pantheon$+$ distance release
\cite{Brout2022}, and the DESI DR2 BAO likelihood \cite{DESIDR2}.  From
Pantheon$+$ we retain the 1580 uncalibrated Hubble-flow light-curve entries satisfying
$z_{\mathrm{HD}}>0.01$ and exclude Cepheid calibrators; the full published
statistical-plus-systematic covariance matrix is restricted to the same
sample.  The DESI DR2 contribution is the published 13-component Gaussian
likelihood, including its full covariance matrix.  The complete likelihood
therefore contains $N=31+1580+13=1624$ entries.

The exact CC inputs and their references are given in
Appendix~\ref{app:CC}. We fix $\Omega_{r0}=9.0\times10^{-5}$ in every
model, independently of the sampled $H_0$; it is not an additional fit
parameter. This convention gives $j_0=1+2\Omega_{r0}=1.00018$ for the
reference flat radiation-plus-matter-plus-$\Lambda$ background.

The CC likelihood neglects off-diagonal correlations between age
estimates and does not add a full stellar-population-synthesis covariance
matrix. The published diagonal errors are used as listed in
Appendix~\ref{app:CC}. Shared stellar-library, metallicity, and population
modelling uncertainties can produce additional covariance
\cite{Moresco2020}; the quoted constraints are conditional on this
diagonal approximation. The present analysis does not establish
robustness to a full CC systematic-covariance treatment.
For $N_{\mathrm{CC}}$ cosmic-chronometer measurements,
\begin{equation}
 \chi^2_{\mathrm{CC}}=
 \sum_{i=1}^{N_{\mathrm{CC}}}
 \frac{\left[H_{\mathrm{obs}}(z_i)-H_{\mathrm{th}}(z_i)\right]^2}
 {\sigma_{H,i}^2}.
\end{equation}
For Pantheon$+$, the luminosity distance for the $i$th light-curve entry is evaluated
using the Hubble-diagram and heliocentric redshifts according to the
convention of the data release,
\begin{equation}
 D_{L,i}=(1+z_{\mathrm{hel},i})D_M(z_{\mathrm{HD},i}).
\end{equation}
The theoretical corrected apparent magnitude and the corresponding
residual are
\begin{align}
 m_{\mathrm{th},i}
 &=5\log_{10}\!\left(\frac{D_{L,i}}{\mathrm{Mpc}}\right)+25+M_B,\\
 \Delta m_i&=m_{b,\mathrm{corr},i}-m_{\mathrm{th},i}.
\end{align}
The supernova contribution is therefore
\begin{equation}
 \chi^2_{\mathrm{SN}}=
 \Delta\bm{m}^{\mathrm T}C_{\mathrm{SN}}^{-1}\Delta\bm{m}.
\end{equation}
Thus, $M_B$ is sampled jointly with $H_0$ rather than being absorbed into
an independently tabulated distance modulus.  For BAO,
\begin{equation}
 \chi^2_{\mathrm{BAO}}=
 \Delta\bm{d}^{\mathrm T}C_{\mathrm{BAO}}^{-1}\Delta\bm{d},
\end{equation}
where $\bm d$ contains the published combinations of $D_M/r_d$, $D_H/r_d$, or $D_V/r_d$.
The exact public-release filenames and sources are listed in
Table~\ref{tab:input_files}; local BAO copies are named
\texttt{desi\_dr2\_mean.txt} and \texttt{desi\_dr2\_cov.txt}.
The supernova row mask is
$(z_{\rm HD}>0.01)\wedge(\texttt{IS\_CALIBRATOR}=0)$; the identical
ordered indices select both covariance axes. Repeated observations are
retained with the published covariance, rather than counted as distinct
supernovae. The DESI ordering follows the mean file: $D_V/r_d$ at
$z=0.295$, followed by $(D_M/r_d,D_H/r_d)$ at
$z=0.510,0.706,0.934,1.321,1.484$, and finally
$(D_H/r_d,D_M/r_d)$ at $z=2.33$.
The total statistic is
\begin{equation}
 \chi^2_{\mathrm{tot}}=
 \chi^2_{\mathrm{CC}}+\chi^2_{\mathrm{SN}}+\chi^2_{\mathrm{BAO}}.
\end{equation}

All three terms are evaluated from the same parameter vector.  Comoving
distances are computed on a deterministic dense redshift grid and
integrated by the cumulative trapezoidal rule. Each distance call uses
$n_{\rm grid}=\max[1200,\lfloor500z_{\max}\rfloor]$ uniformly spaced
nodes, followed by linear interpolation. For the present SN and BAO
samples this is 1200 nodes. At each updated joint optimum, replacing
this grid by 9600 nodes changes the total $\chi^2$ by less than
$0.003$, and the SN comoving distances by less than
$2.6\times10^{-5}$ fractionally. Checks against adaptive quadrature
at $z=0.01,0.1,0.5,1,2.33$ agree with the dense-grid distances to
better than $2\times10^{-9}$ fractionally. These are numerical
accuracy checks at the fitted optima, not bounds over the entire prior.
Differences of a few $10^{-3}$ in reported $\chi^2$ should therefore
not be assigned physical significance.  The Pantheon$+$
covariance matrix is factorized once by a Cholesky decomposition; each
likelihood call then uses triangular solves rather than forming an
explicit inverse.  The same procedure is applied to the DESI covariance.
As a numerical consistency check, the code verifies that the GUP model at
$\betastar=0$ and the CPL model at $(w_0,w_a)=(-1,0)$ reproduce the
\LCDM{} expansion rate and all three likelihood contributions to machine
precision. These nested-limit checks are separate from the integration-accuracy tests above.

We adopt the uniform priors listed in Table~\ref{tab:priors}.  For the GUP
model, samples must additionally satisfy $\Omega_{\Lambda0}>0$,
$E^2(z)>0$, and $\max|\beta p_a^2|<0.1$ over the fitted interval
$0\leq z\leq2.33$.
For later use, define the expansion measure
\begin{equation}
 \eta_{\max}\equiv
 \max_{0\leq z\leq2.33}
 \left[\frac12|\betastar|(1+z)^{-4}X(z)\right].
 \label{eq:eta_max}
\end{equation}
For the strictly order-by-order model of Eq.~\eqref{eq:E2_strict}, the
validity measure is evaluated consistently at leading order by replacing
$X(z)$ with $Y(z)$:
\begin{equation}
 \eta_{\max}^{\rm strict}\equiv
 \max_{0\leq z\leq2.33}
 \left[\frac12|\betastar|(1+z)^{-4}Y(z)\right].
 \label{eq:eta_max_strict}
\end{equation}
For $0<a\leq1$ and positive matter and constant densities,
\begin{equation}
 \frac{d(a^4X)}{da}=\Omega_{m0}+4\Omega_{\Lambda0}a^3>0.
\end{equation}
The same argument applies to $Y$ because
$1-\Omega_{r0}-\Omega_{m0}>0$ under the adopted priors. Hence
\begin{equation}
 \eta_{\max}=\frac{|\betastar|X_0}{2},\qquad
 \eta_{\max}^{\rm strict}=\frac{|\betastar|}{2}.
 \label{eq:eta_analytic}
\end{equation}
On the baseline branch,
$\eta_{\max}=|\sqrt{1+4\betastar}-1|/4$.
For $0<\eta_c<1/4$, the condition $\eta_{\max}<\eta_c$ is equivalent to
\begin{equation}
 -2\eta_c+4\eta_c^2<\betastar<2\eta_c+4\eta_c^2,
 \label{eq:cut_prior}
\end{equation}
intersected with the box prior in Table~\ref{tab:priors}. In the strict
model it is $-2\eta_c<\betastar<2\eta_c$. Thus the negative prior
boundaries are $-0.1600,-0.1275,-0.0900$ for baseline cutoffs
$0.10,0.075,0.05$, and $-0.2000,-0.1500,-0.1000$ for the strict model.
These are imposed validity boundaries, not observational confidence
limits. In particular, the tight-cut lower $95\%$ limits in
Table~\ref{tab:robustness} lie close to these boundaries.

Using $X(z)=Y(z)-\betastar$ inside this criterion would introduce terms
beyond the retained first order.
The broad validity choice is $\eta_{\max}<0.10$ for the exact-root
baseline and $\eta_{\max}^{\rm strict}<0.10$ for the strict model.
More restrictive values are
examined in Sec.~\ref{sec:robustness}; they are imposed on the saved
broad-prior chain as nested hard cuts, leaving the likelihood and all
other priors unchanged.
\begin{table}[!htbp]
\centering
\caption{Uniform box priors before the additional validity cuts. The effective negative $\betastar$ boundary depends on the normalization and cutoff through Eq.~\eqref{eq:cut_prior}.}
\label{tab:priors}

\small
\setlength{\tabcolsep}{4pt}
\renewcommand{\arraystretch}{1.05}

\begin{tabular}{@{}lc@{}}
\hline\hline
\noalign{\vskip 1.5mm}
Parameter & Prior\\
\noalign{\vskip 0.8mm}
\hline
\noalign{\vskip 0.5mm}
$H_0\,[\mathrm{km\,s^{-1}\,Mpc^{-1}}]$ & $[50,90]$\\
$\Omega_{m0}$                           & $[0.1,0.5]$\\
$\beta_{*}$ (GUP only)                  & $[-0.25,0.1]$\\
$r_d\,[\mathrm{Mpc}]$                   & $[120,170]$\\
$M_B$                                   & $[-20.5,-18.0]$\\
$w_0$ (CPL only)                        & $[-2,0]$\\
$w_a$ (CPL only)                        & $[-3,3]$\\
\noalign{\vskip 0.5mm}
\hline\hline
\end{tabular}

\end{table}
Sampling quality is assessed using the integrated autocorrelation time,
the corresponding effective sample size, and the stability of the
retained chains, as reported in Sec.~\ref{sec:results}.

\section{Reference models and statistical comparison}\label{sec:comparison}

The GUP background and the two reference models are fitted to exactly the
same datasets.  For flat \LCDM,
\begin{equation}
 E_{\Lambda\mathrm{CDM}}^2(z)
 =\Omega_{r0}(1+z)^4+\Omega_{m0}(1+z)^3
 +1-\Omega_{r0}-\Omega_{m0}.
\end{equation}
For CPL,
\begin{align}
 E_{\mathrm{CPL}}^2(z)={}&\Omega_{r0}(1+z)^4
 +\Omega_{m0}(1+z)^3\nonumber\\
 &+\Omega_{\mathrm{DE},0}
 (1+z)^{3(1+w_0+w_a)}
 \exp\!\left[-\frac{3w_az}{1+z}\right],
\end{align}
where $\Omega_{\mathrm{DE},0}=1-\Omega_{r0}-\Omega_{m0}$.

In addition to $\chi^2_{\min}$, we use the Akaike and Bayesian information
criteria \cite{Akaike1974,Schwarz1978},
\begin{equation}
 \mathrm{AIC}=\chi^2_{\min}+2k,
 \qquad
 \mathrm{BIC}=\chi^2_{\min}+k\ln N,
\end{equation}
where $k$ is the number of independently fitted parameters and $N$ is the number of data points.  For each information criterion, differences are quoted relative to the model with the smallest value:
\begin{equation}
 \Delta\mathrm{IC}=\mathrm{IC}_{\mathrm{model}}-
 \mathrm{IC}_{\min}.
\end{equation}
\section{Numerical results}\label{sec:results}

\begin{figure*}[!t]
\centering
\includegraphics[width=.88\textwidth,height=0.76\textheight,keepaspectratio]{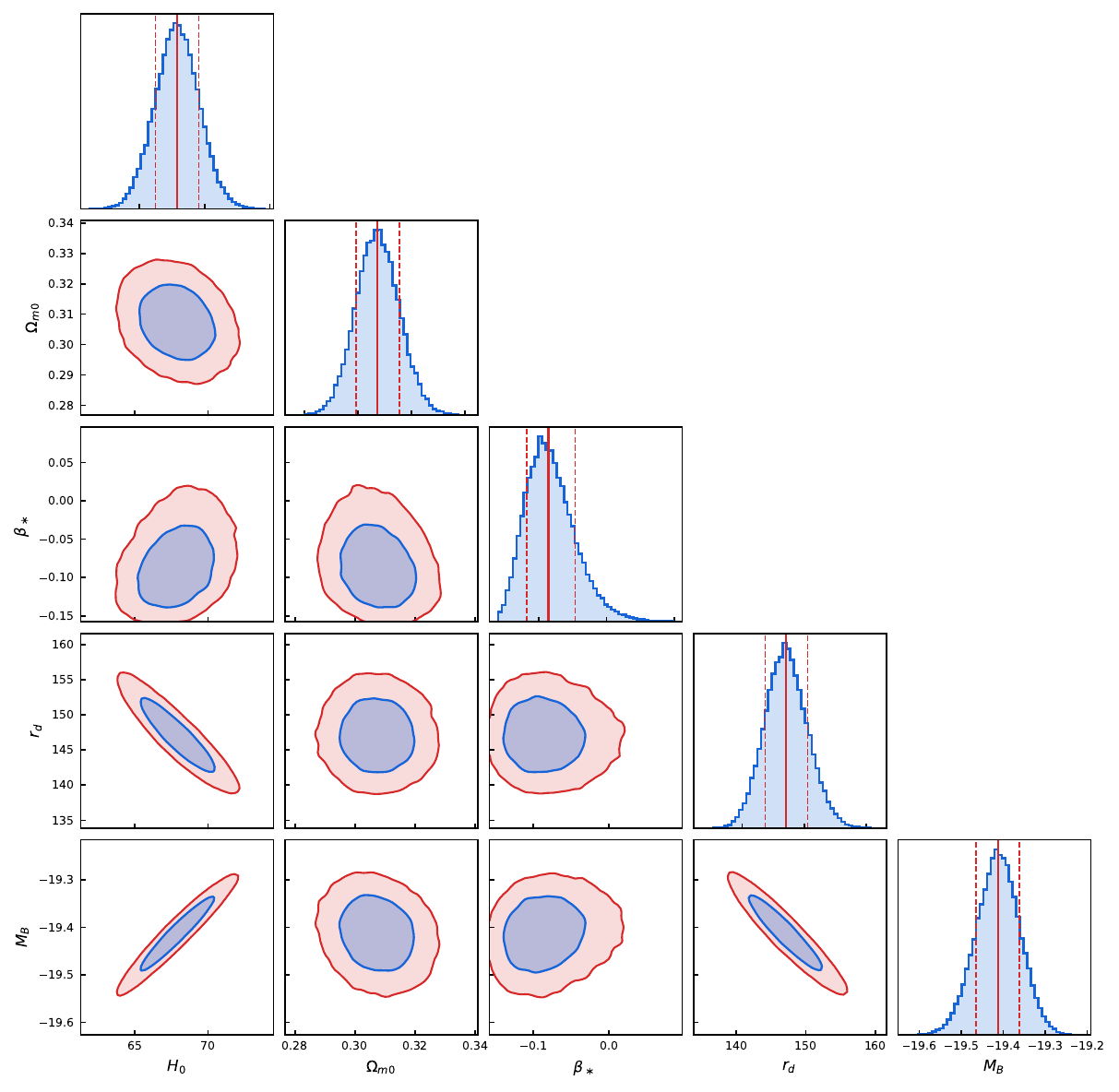}
\caption{Marginalized one- and two-dimensional posterior distributions for
the GUP background model obtained from the combined
CC+Pantheon$+$+DESI DR2 likelihood.  The quoted one-dimensional intervals
are equal-tail $68\%$ credible intervals.  The blue and red contours
enclose the $68\%$ and $95\%$ posterior probabilities, respectively,
while the red vertical lines in the diagonal panels indicate the median
and equal-tail $68\%$ credible limits.}
\label{fig:gup_posterior}
\end{figure*}

The posterior was sampled with the affine-invariant ensemble method
implemented in \textsc{emcee} \cite{ForemanMackey2013}, using 32 walkers.
After burn-in, the \LCDM{} and GUP runs each contain $256\,000$ retained
samples.  The CPL run was extended to $70\,000$ retained iterations per
walker.  Integrated-autocorrelation-time estimates give minimum effective
sample sizes of approximately $4890$, $4040$, and $2180$ for \LCDM{}, GUP,
and CPL, respectively.  For the CPL run, the largest integrated
autocorrelation time is approximately $1029$ iterations, and the maximum
shift between the two halves of the retained chain is $0.040$ posterior
standard deviations.  These diagnostics describe the original retained ensembles; walkers in
an affine-invariant ensemble are coupled and should not be interpreted
as independent chains for a conventional Gelman--Rubin test.
The strict run also used 32 walkers, 12000 steps, and a 4000-step burn-in;
its largest autocorrelation estimate is about 68 steps.

The archived CC table used $\sigma_H=34$ at $z=0.47$. We correct this
entry to $49.6~\mathrm{km\,s^{-1}\,Mpc^{-1}}$, consistent with the updated
CC compilation and the combined statistical and systematic uncertainty
of Ratsimbazafy et al.~\cite{Ratsimbazafy2017}. To update the posterior
without rerunning the ensembles, each archived sample is assigned
\begin{equation}
 w_i\propto\exp\!\left[-\frac{[H_i(0.47)-89]^2}{2}
 \left(\frac{1}{49.6^2}-\frac{1}{34^2}\right)\right],
 \label{eq:importance_update}
\end{equation}
with $H$ expressed in $\mathrm{km\,s^{-1}\,Mpc^{-1}}$.
All intervals below use these normalized weights. The efficiency
$(\sum_iw_i)^2/(N_{\rm samp}\sum_iw_i^2)$ exceeds $0.99999$ for all
four ensembles; the changes in parameter summaries are much smaller than their quoted statistical uncertainties. This is an importance update, not an independent MCMC run.
Filtered probabilities in Table~\ref{tab:robustness} are weighted
posterior masses. For the baseline and strict broad samples, the
estimated Monte Carlo standard errors of the $\betastar$ means are
$0.00059$ and $0.00080$. Across the tighter selections they are below
$0.00060$, with approximate mean effective sample sizes above 4200.
These estimates use ratio-estimator influence functions on the original
step-by-walker grid, preserving time ordering before filtering; the
two-half conditional-mean shifts are below $0.022$ posterior standard
deviations. These diagnostics assess conditional means, not the error
of every tail quantile.

The joint minima are recomputed with the corrected CC entry using
Nelder--Mead initialized at the archived differential-evolution optima.
The archived global search used seed 1731, population factor 12, and
250 maximum generations. The updated local searches use at most 3500
iterations, with parameter and objective tolerances $10^{-8}$.
All four local searches terminate successfully. Tables~\ref{tab:optima}
and \ref{tab:chi_components} give the resulting joint optima and
likelihood decomposition; these points are distinct from posterior
medians and from the best samples in a finite chain.

\begin{table*}[!t]
\centering
\caption{Joint best-fit parameter vectors after the CC-error correction. $H_0$ is in $\mathrm{km\,s^{-1}\,Mpc^{-1}}$ and $r_d$ in Mpc. A dash denotes a parameter absent from that model. Values are rounded from the optimizer output, not posterior medians.}
\label{tab:optima}
\small
\begin{tabular}{lrrrrrrr}
\hline\hline
Model & $H_0$ & $\Omega_{m0}$ & $\betastar$ & $w_0$ & $w_a$ & $r_d$ & $M_B$\\
\hline
$\Lambda$CDM & 68.827415 & 0.304411 & --- & --- & --- & 146.718023 & -19.397180\\
GUP & 67.847736 & 0.307530 & -0.095145 & --- & --- & 146.919220 & -19.411780\\
GUP (strict) & 67.848558 & 0.308397 & -0.129592 & --- & --- & 146.919748 & -19.411715\\
CPL & 67.850909 & 0.304971 & --- & -0.891243 & -0.201496 & 146.893275 & -19.413091\\
\hline\hline
\end{tabular}
\end{table*}
\begin{table*}[!t]
\centering
\small
\caption{Likelihood decomposition at the same joint optima as Table~\ref{tab:optima}. Component minima are not optimized separately.}
\label{tab:chi_components}
\begin{tabular}{lrrrr}
\hline\hline
Model & $\chi^2_{\rm CC}$ & $\chi^2_{\rm SN}$ & $\chi^2_{\rm BAO}$ & $\chi^2_{\rm tot}$\\
\hline
$\Lambda$CDM & 14.566 & 1388.867 & 10.976 & 1414.410\\
GUP & 14.737 & 1386.094 & 8.892 & 1409.722\\
GUP (strict) & 14.740 & 1386.098 & 8.886 & 1409.724\\
CPL & 14.749 & 1386.045 & 8.696 & 1409.490\\
\hline\hline
\end{tabular}
\end{table*}

The marginalized constraint on the deformed-minisuperspace coefficient is
\begin{equation}
 \betastar=-0.086^{+0.040}_{-0.032}
 \qquad (68\%\ \text{credible interval}),
 \label{eq:beta_constraint}
\end{equation}
while the equal-tail $95\%$ credible interval is
\begin{equation}
 -0.142<\betastar<0.005.
\end{equation}
The best-fit value is $\betastar=-0.095$.  Hence, although the
one-dimensional posterior is concentrated at negative values, the
undeformed limit $\betastar=0$ remains inside the $95\%$ interval.  The
negative tail is limited in part by the imposed first-order validity
condition.  The quoted interval must therefore be interpreted conditional
on that perturbative cutoff and should not be extrapolated into the
nonperturbative region.
The
principal marginalized constraints are summarized in
Table~\ref{tab:parameters}.

\begin{table*}[!t]
\centering
\caption{Marginalized medians and $68\%$ credible intervals for the
combined CC+Pantheon$+$+DESI DR2 likelihood.}
\label{tab:parameters}

\small
\setlength{\tabcolsep}{5pt}
\renewcommand{\arraystretch}{1.08}

\begin{tabular}{lccc}\hline\hline Parameter & $\Lambda$CDM & GUP & CPL\\\hline
$H_0$ & $68.79^{+1.62}_{-1.63}$ & $67.88^{+1.68}_{-1.68}$ & $67.82^{+1.64}_{-1.63}$\\
$\Omega_{m0}$ & $0.3046^{+0.0079}_{-0.0078}$ & $0.3072^{+0.0082}_{-0.0079}$ & $0.3023^{+0.0160}_{-0.0269}$\\
$\betastar$ & $0\;(\mathrm{fixed})$ & $-0.086^{+0.040}_{-0.032}$ & $0\;(\mathrm{fixed})$\\
$w_0$ & $-1\;(\mathrm{fixed})$ & $-1\;(\mathrm{fixed})$ & $-0.890^{+0.062}_{-0.057}$\\
$w_a$ & $0\;(\mathrm{fixed})$ & $0\;(\mathrm{fixed})$ & $-0.135^{+0.518}_{-0.458}$\\
$r_d\,[\mathrm{Mpc}]$ & $146.81^{+3.49}_{-3.34}$ & $147.05^{+3.57}_{-3.33}$ & $146.95^{+3.43}_{-3.28}$\\
$M_B$ & $-19.399^{+0.050}_{-0.052}$ & $-19.413^{+0.051}_{-0.053}$ & $-19.415^{+0.050}_{-0.051}$\\
\hline\hline\end{tabular}

\end{table*}

\begin{table*}[!t]
\centering
\caption{Statistical comparison of the $\Lambda$CDM, GUP, and CPL
models for the combined CC+Pantheon$+$+DESI DR2 data. The differences
$\Delta\mathrm{AIC}$ and $\Delta\mathrm{BIC}$ are calculated relative
to the model having the smallest value of the corresponding
information criterion.}
\label{tab:model_comparison}
\small
\setlength{\tabcolsep}{6pt}
\begin{tabular}{lcccccc}
\hline\hline
Model & $k$ & $\chi^2_{\min}$ & $\mathrm{AIC}$
& $\Delta\mathrm{AIC}$ & $\mathrm{BIC}$
& $\Delta\mathrm{BIC}$ \\
\hline
$\Lambda$CDM & 4 & 1414.410 & 1422.410 & 2.687
             & 1443.980 & 0.000 \\
GUP          & 5 & 1409.722 & 1419.722 & 0.000
             & 1446.686 & 2.705 \\
CPL          & 6 & 1409.490 & 1421.490 & 1.768
             & 1453.846 & 9.866 \\
\hline\hline
\end{tabular}
\end{table*}

The deformed-minisuperspace and CPL extensions yield lower minimum chi-square values than 
$\Lambda$CDM, with improvements of
$\Delta\chi^2=-4.687$ and $-4.919$, respectively. These improvements
must, however, be assessed together with the additional model
parameters. According to the AIC, the deformed-minisuperspace model gives the smallest
value, while the CPL and $\Lambda$CDM models have
$\Delta\mathrm{AIC}=1.768$ and $2.687$, respectively. Thus, the AIC
provides only a mild preference for the deformation and does not
establish decisive evidence against the alternatives. In contrast,
the BIC favors the more economical $\Lambda$CDM model. The deformed model
has $\Delta\mathrm{BIC}=2.705$, whereas the CPL parametrization has
$\Delta\mathrm{BIC}=9.866$. The data therefore permit an improved
background fit in the extended models, but the statistical evidence
is not sufficient to establish a departure from the undeformed
background.  In view of Sec.~\ref{sec:decoupling}, this comparison cannot be promoted to
a test of fundamental Planck-scale physics.

\section{Perturbative robustness and background diagnostics}

\begin{figure}[!htbp]
\centering
\includegraphics[width=\linewidth]{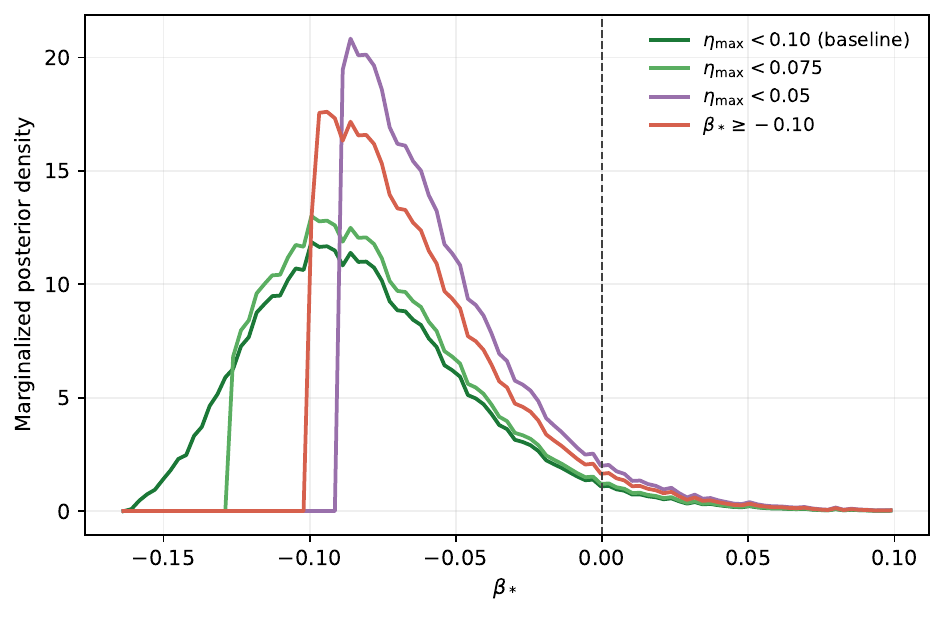}
\caption{Marginalized $\betastar$ distributions obtained from the saved
broad-prior chain after nested perturbative and prior cuts.  The
sensitivity of the negative tail demonstrates why the baseline interval
must be quoted together with its validity convention.}
\label{fig:robustness}
\end{figure}

\begin{figure*}[!t]
\centering
\includegraphics[width=.98\textwidth,height=0.76\textheight,keepaspectratio]{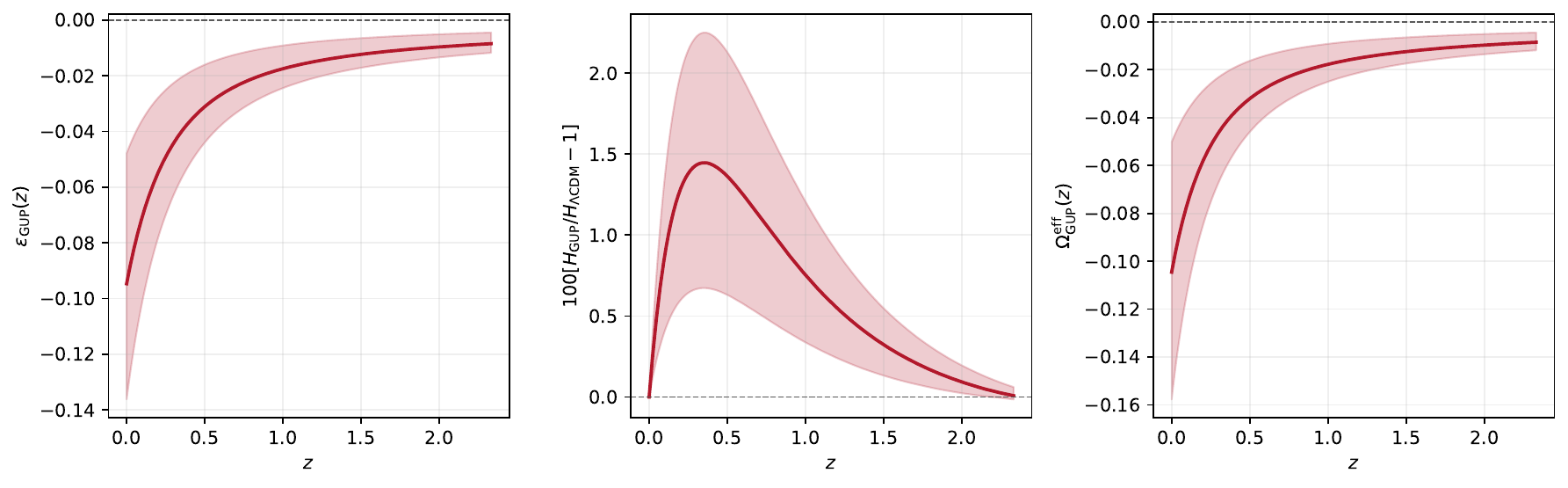}
\caption{Posterior medians and 68\% credible bands for the fractional
GUP correction (left), the fractional shift in $H(z)$ relative to a
flat-\LCDM{} background matched in $H_0$, $\Omega_{m0}$, and $\Omega_{r0}$ in percent (center), and
the effective GUP fraction (right).}
\label{fig:background_diagnostics}
\end{figure*}
\label{sec:robustness}

The baseline posterior is not independent of the quantitative meaning
assigned to ``first order.''  This issue is especially relevant on the
negative side, where the normalization branch makes $X_0$ increase as
$\betastar$ approaches its algebraic lower bound.  We therefore repeat
the posterior summary after imposing the nested cuts
$\eta_{\max}<0.075$ and $\eta_{\max}<0.05$.  We also impose
$\betastar\geq-0.10$ as a direct prior-sensitivity test.  Because these
are nested hard cuts inside the support of the original uniform-prior
chain, the restricted distributions follow by exact filtering of the
saved samples with the importance weights of Eq.~\eqref{eq:importance_update}; no Gaussian approximation is introduced.

\begin{table*}[!t]
\centering
\caption{Sensitivity of the marginalized GUP constraint to the
normalization prescription and to nested perturbative and prior cuts.
The retained fraction is the weighted posterior mass relative to the corresponding broad-prior chain. The negative support boundaries are given in Eq.~\eqref{eq:cut_prior}.}
\label{tab:robustness}
\small
\setlength{\tabcolsep}{6pt}
\begin{tabular}{lccc}\hline\hline Selection & $\betastar$ (68\%) & $\betastar$ (95\%) & Posterior mass\\\hline
$\eta_{\max}<0.10$ (baseline) & $-0.086^{+0.040}_{-0.032}$ & $[-0.142,\,0.005]$ & 1.000\\
$\eta_{\max}<0.075$ (baseline) & $-0.082^{+0.039}_{-0.028}$ & $[-0.124,\,0.007]$ & 0.912\\
$\eta_{\max}<0.05$ (baseline) & $-0.063^{+0.034}_{-0.019}$ & $[-0.089,\,0.020]$ & 0.547\\
$\betastar\geq-0.10$ (baseline) & $-0.069^{+0.036}_{-0.021}$ & $[-0.099,\,0.015]$ & 0.663\\
$\eta_{\max}^{\rm strict}<0.10$ & $-0.119^{+0.055}_{-0.048}$ & $[-0.194,\,-0.010]$ & 1.000\\
$\eta_{\max}^{\rm strict}<0.075$ & $-0.100^{+0.047}_{-0.034}$ & $[-0.147,\,-0.002]$ & 0.728\\
$\eta_{\max}^{\rm strict}<0.05$ & $-0.069^{+0.038}_{-0.022}$ & $[-0.099,\,0.013]$ & 0.361\\
\hline\hline\end{tabular}
\end{table*}

Table~\ref{tab:robustness} shows that the conditional posterior continues to favor
negative values under the more conservative selections, but the median
and lower credible boundary move appreciably.  In particular, the
$\eta_{\max}<0.05$ posterior gives
$\betastar=-0.063^{+0.034}_{-0.019}$ and includes $\betastar=0$ comfortably
at $95\%$ credibility.  The baseline result is therefore not a
cutoff-independent measurement.  The median baseline expansion measure
is $\eta_{\max}=0.047^{+0.021}_{-0.023}$.  Its 95th and 97.5th percentiles
are approximately $0.080$ and $0.086$, respectively.  Equation~\eqref{eq:eta_analytic} proves that the maximum occurs at $z=0$ throughout the fitted interval;
the high-redshift BAO points do not drive the perturbative boundary.

\begin{figure*}[!t]
\centering
\includegraphics[width=.98\textwidth,height=0.76\textheight,keepaspectratio]{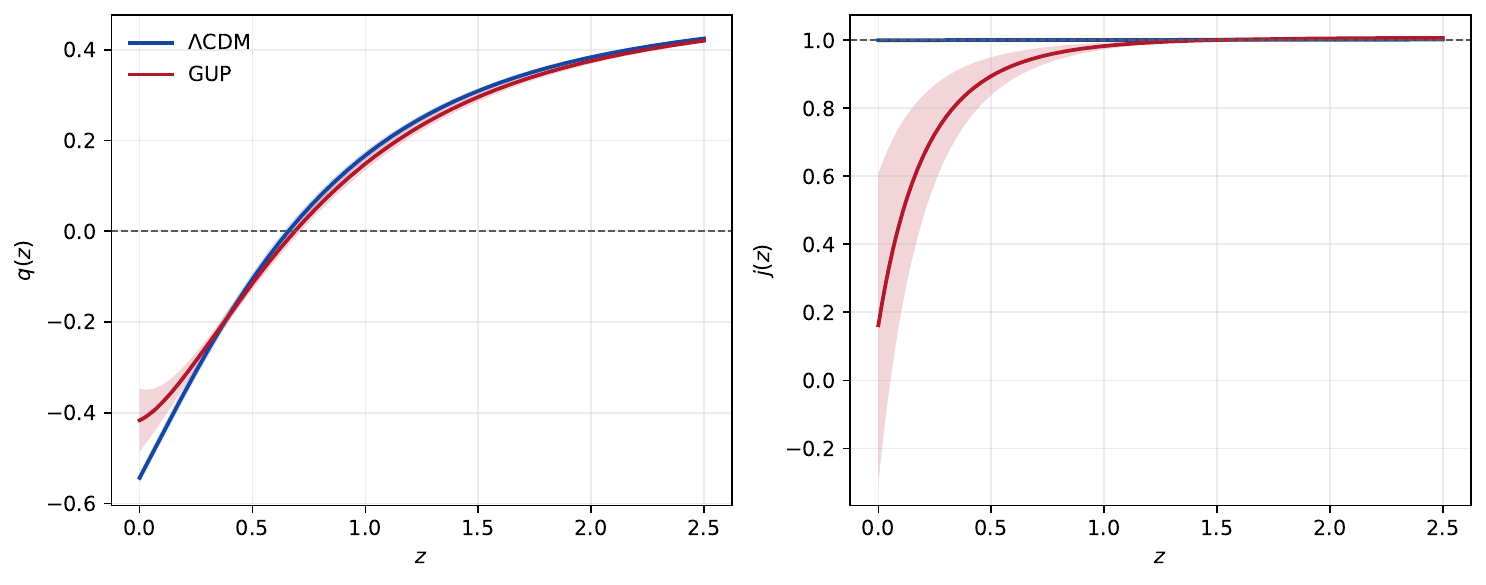}
\caption{Posterior medians and $68\%$ credible bands of the
deceleration parameter (left) and jerk parameter (right) for the GUP
model and \LCDM{}. Dashed lines indicate $q=0$ and the \LCDM{}
late-time value $j=1$, respectively.}
\label{fig:cosmography}
\end{figure*}

The independent chain based on Eq.~\eqref{eq:E2_strict} gives
$\betastar=-0.119^{+0.055}_{-0.048}$ and
$\chi^2_{\min}=1409.724$, compared with $1409.722$ for the exact-root
baseline.  The two prescriptions therefore attain indistinguishable
best-fit quality while shifting the inferred deformation.  Under the
broad $\eta_{\max}<0.10$ criterion, the strict-normalization interval
excludes zero at nominal $95\%$ credibility.  This conclusion disappears
for $\eta_{\max}<0.05$, for which the interval is
$[-0.099,\,0.013]$ and only $36.1\%$ of the strict-chain samples are
retained.  Thus neither the numerical constraint nor an apparent
exclusion of the undeformed limit is robust to the combined
normalization and cutoff choices.

Figure~\ref{fig:background_diagnostics} propagates the broad-prior
posterior to three related measures.  The fractional deformation
$\epsilon_{\rm GUP}$ is negative for the median fit and becomes less
important toward the matter-dominated past.  The middle panel compares
the GUP rate with a flat \LCDM{} background having the same $H_0$, $\Omega_{m0}$, and fixed $\Omega_{r0}$; this matched-parameter comparison isolates the deformation
from shifts in the fitted matter density.  The final panel shows the
effective fractional contribution defined in
Eq.~\eqref{eq:OmegaGUPeff}.  None of these quantities is an independent
observable, but together they display where the fitted background differs
from its undeformed counterpart.

\section{Posterior cosmography and acceleration history}

\label{sec:cosmography}

The fitted background can be characterized without introducing a
perturbation prescription through the deceleration and jerk parameters,
\begin{align}
 q(z)&=-1+\frac{1+z}{E(z)}\frac{dE(z)}{dz},
 \label{eq:qz}\\
 j(z)&=q(z)\left[2q(z)+1\right]+(1+z)\frac{dq(z)}{dz}.
 \label{eq:jz}
\end{align}
The acceleration-transition redshift is defined by
$q(z_{\mathrm{tr}})=0$.  We evaluate these expressions for every
retained posterior sample, thereby preserving parameter correlations
instead of evaluating them only at a marginalized median point. For
$F(z)=E^2(z)$, the equivalent analytic relations
\begin{align}
 q(z)&=-1+\frac{(1+z)F'}{2F},\\
 j(z)&=1-\frac{(1+z)F'}{F}+\frac{(1+z)^2F''}{2F}
\end{align}
avoid finite-difference amplification in the jerk. Weighted summaries
use the full saved samples; plotted bands use 30000 weighted resampled
draws with a fixed seed, as documented in the accompanying code.

We also define the additive GUP contribution as a fraction of the total
normalized expansion rate,
\begin{equation}
 \Omega_{\mathrm{GUP}}^{\mathrm{eff}}(z)
 =\frac{\betastar(1+z)^{-4}X^2(z)}{E^2(z)}.
 \label{eq:OmegaGUPeff}
\end{equation}
This is a bookkeeping quantity for the modified background equation,
not the density parameter of an independently conserved fluid.  At
$z=0$ it reduces to
$\Omega_{\mathrm{GUP},0}^{\mathrm{eff}}=\betastar X_0^2$.

\begin{table}[!htbp]
\centering
\caption{Posterior-derived present-day kinematic quantities and
transition redshift. The entries are medians and $68\%$ equal-tail
credible intervals.}
\label{tab:cosmography}

\small
\setlength{\tabcolsep}{4pt}
\renewcommand{\arraystretch}{1.08}

\begin{tabular}{lcc}\hline\hline Quantity & $\Lambda$CDM & GUP\\\hline
$q_0$ & $-0.543^{+0.012}_{-0.012}$ & $-0.416^{+0.070}_{-0.069}$\\
$j_0$ & $1.00018$ & $0.162^{+0.445}_{-0.461}$\\
$z_{\rm tr}$ & $0.658^{+0.021}_{-0.020}$ & $0.688^{+0.027}_{-0.026}$\\
$\Omega_{\rm GUP,0}^{\rm eff}$ & $0$ & $-0.105^{+0.054}_{-0.053}$\\
\hline\hline\end{tabular}

\end{table}

The GUP posterior yields
$q_0=-0.416^{+0.070}_{-0.069}$,
$z_{\mathrm{tr}}=0.688^{+0.027}_{-0.026}$, and
$j_0=0.162^{+0.445}_{-0.461}$, supporting present acceleration. These derived summaries do not supply an independent model-selection test against \LCDM{}. The negative
$\Omega_{\mathrm{GUP},0}^{\mathrm{eff}}$ reduces $E^2$ relative to $X$;
however, because $\betastar=0$ remains allowed at $95\%$ credibility,
these results do not constitute evidence for a GUP deformation.

\section{Discussion and conclusions}\label{sec:conclusions}

We have constrained a quadratic deformation of the reduced Poisson algebra in an FLRW minisuperspace model. This construction is a phenomenological ansatz introduced after imposing homogeneity, which removes local gravitational degrees of freedom. It is not derived from a covariant ultraviolet theory, does not provide an invariant connection between $\betastar$ and a microscopic minimum length, and does not explain how Planck-scale physics could avoid low-energy decoupling. Therefore, our analysis is not a direct observational test of quantum gravity, but a parameter-estimation study of a particular reduced cosmological model.

The modified Friedmann equation, Eq.~\eqref{eq:E2}, defines a changed homogeneous expansion history and its associated late-time distance relations. However, the minisuperspace deformation does not provide equations for cosmological perturbations. We therefore restrict the analysis to background data and do not use CMB anisotropies, structure-growth measurements, weak lensing, or the matter power spectrum. The BAO sound horizon is treated as a nuisance parameter rather than as a prediction of the model.

An important limitation is the normalization ambiguity of the truncated model. In the baseline treatment, the closure relation is solved algebraically using Eq.~\eqref{eq:X0}, whereas the strict treatment applies the expansion order by order through Eq.~\eqref{eq:E2_strict}. The different posterior distributions obtained from these approaches show that formally higher-order terms are comparable to the statistical uncertainty. In addition, the relative correction remains constant during radiation domination and decreases toward the past during matter domination. Thus, a larger background density does not necessarily increase the fractional deformation.

For the baseline normalization, the combined CC+Pantheon$+$+DESI DR2 analysis gives
$\betastar=-0.086^{+0.040}_{-0.032}$, with a $95\%$ credibility interval of $[-0.142,\,0.005]$. The model reduces the minimum chi-square relative to flat \LCDM{} by $4.687$. However, the AIC provides only mild support for the extra parameter, while the BIC favors \LCDM{}. Since the undeformed value remains allowed at $95\%$ credibility, the result should be interpreted as a null constraint rather than evidence for a nonzero deformation. Moreover, the negative posterior median has the opposite sign to the conventional positive-coefficient minimum-length GUP and cannot be regarded as support for that scenario.

The robustness tests lead to the same conclusion. Tightening the baseline cutoff from $\eta_{\max}<0.10$ to $\eta_{\max}<0.05$ reduces the negative posterior tail. Under the broad cutoff, the strict order-by-order treatment gives $\betastar=-0.119^{+0.055}_{-0.048}$; with the conservative condition $\eta_{\max}^{\rm strict}<0.05$, the $95\%$ interval becomes $[-0.099,\,0.013]$, which again includes zero. The inferred value of the deformation parameter, and any apparent exclusion of the undeformed limit, are therefore not robust under reasonable normalization and truncation choices.

A fundamental interpretation would require much more than a modified background expansion history. A complete theory would need a covariant formulation, a clear identification of the ultraviolet scale and its decoupling behavior, a closed constraint algebra, freedom from fiducial-volume ambiguities, and derived equations for scalar, vector, and tensor perturbations. Only then could CMB, lensing, structure-growth data, and a model-derived sound horizon be used consistently. Until such a completion is available, the present results should be interpreted only as conditional constraints on the phenomenological minisuperspace model considered here.

\appendix
\section{Adopted cosmic-chronometer inputs}\label{app:CC}
Table~\ref{tab:cc_inputs} fixes the exact rounded numerical inputs used
by the likelihood. They are retained for reproducibility rather than
silently replaced by newer measurements. Radial BAO determinations of
$H(z)$ are excluded. At $z=0.47$, the corrected total error is 49.6;
the older value 34 is used only to construct the importance update in
Eq.~\eqref{eq:importance_update}. Statistical and systematic errors
for this point are combined in quadrature in the updated compilation.
Other CC modelling correlations are subject to the diagonal-likelihood
limitation stated in Sec.~\ref{sec:data}.
\begin{table*}[!t]
\centering
\caption{The 31 adopted CC measurements. Both $H$ and $\sigma_H$ are in $\mathrm{km\,s^{-1}\,Mpc^{-1}}$. References identify the original measurement studies.}
\label{tab:cc_inputs}
\small
\begin{tabular}{rrrr@{\hspace{12mm}}rrrr}
\hline\hline
$z$ & $H$ & $\sigma_H$ & Ref. & $z$ & $H$ & $\sigma_H$ & Ref.\\\hline
0.07 & 69 & 19.6 & \cite{Zhang2014} & 0.4783 & 80.9 & 9 & \cite{Moresco2016}\\
0.09 & 69 & 12 & \cite{Simon2005} & 0.48 & 97 & 62 & \cite{Stern2010}\\
0.12 & 68.6 & 26.2 & \cite{Zhang2014} & 0.593 & 104 & 13 & \cite{Moresco2012}\\
0.17 & 83 & 8 & \cite{Simon2005} & 0.68 & 92 & 8 & \cite{Moresco2012}\\
0.179 & 75 & 4 & \cite{Moresco2012} & 0.781 & 105 & 12 & \cite{Moresco2012}\\
0.199 & 75 & 5 & \cite{Moresco2012} & 0.875 & 125 & 17 & \cite{Moresco2012}\\
0.2 & 72.9 & 29.6 & \cite{Zhang2014} & 0.88 & 90 & 40 & \cite{Stern2010}\\
0.27 & 77 & 14 & \cite{Simon2005} & 0.9 & 117 & 23 & \cite{Simon2005}\\
0.28 & 88.8 & 36.6 & \cite{Zhang2014} & 1.037 & 154 & 20 & \cite{Moresco2012}\\
0.352 & 83 & 14 & \cite{Moresco2012} & 1.3 & 168 & 17 & \cite{Simon2005}\\
0.3802 & 83 & 13.5 & \cite{Moresco2016} & 1.363 & 160 & 33.6 & \cite{Moresco2015}\\
0.4 & 95 & 17 & \cite{Simon2005} & 1.43 & 177 & 18 & \cite{Simon2005}\\
0.4004 & 77 & 10.2 & \cite{Moresco2016} & 1.53 & 140 & 14 & \cite{Simon2005}\\
0.4247 & 87.1 & 11.2 & \cite{Moresco2016} & 1.75 & 202 & 40 & \cite{Simon2005}\\
0.4497 & 92.8 & 12.9 & \cite{Moresco2016} & 1.965 & 186.5 & 50.4 & \cite{Moresco2015}\\
0.47 & 89 & 49.6 & \cite{Ratsimbazafy2017} & & & & \\
\hline\hline
\end{tabular}
\end{table*}

\section*{Data Availability Statement}

This study reanalyzes publicly released observations; no new
observational measurements were collected. The Pantheon$+$ data and
covariance are available at
\url{https://github.com/PantheonPlusSH0ES/DataRelease}, and the DESI DR2
BAO mean vector and covariance at
\url{https://github.com/CobayaSampler/bao_data/tree/master/desi_bao_dr2}.
The exact adopted CC table is given in Appendix~\ref{app:CC} and as a
machine-readable file in the accompanying reproducibility package.
That package contains the likelihood code, archived chains, input-data
checksums, importance weights, revised numerical summaries, and figure
scripts. 

\section*{Declarations}

\textbf{Funding:} No funding was received for this work.

\textbf{Competing Interests:} The author declares that there are no competing interests.

\end{document}